\documentclass[lettersize,journal]{IEEEtran}

\usepackage{stmaryrd}
\usepackage{amsmath}
\usepackage{amssymb}
\usepackage{enumitem}
\usepackage{psfrag}
\usepackage{graphicx}
\usepackage{epstopdf}
\usepackage{multirow}
\usepackage{booktabs}
\usepackage{siunitx}
\usepackage{cite}
\usepackage{algorithmic}
\usepackage{algorithm}
\usepackage{mathtools}
\usepackage{hyperref}
\usepackage{filecontents}
\usepackage{bm}

\usepackage{float}

\usepackage{caption}
\usepackage{subcaption}
\usepackage[inkscapelatex=false]{svg}
\ifCLASSOPTIONcompsoc
\else
\fi
\usepackage{comment}

\begin{document}

\title{
A Universal Neural Receiver that Learns at the Speed of Wireless
}
\author{Lingjia Liu, Lizhong Zheng, Yang Yi, and Robert Calderbank
\thanks{
L. Liu and Y. Yi are with Wireless@Virginia Tech, ECE Department at Virginia Tech. L. Zheng is with the EECS Department at the Massachusetts Institute of Technology (MIT), and R. Calderbank is with the ECE Department at Duke University. 
The corresponding author is L. Liu (ljliu@ieee.org).
}
}
\maketitle

\begin{abstract}
Today we design wireless networks using mathematical models that govern communication in different propagation environments. We rely on measurement campaigns to deliver parametrized propagation models, and on the 3GPP standards process to optimize model-based performance, but as wireless networks become more complex this model-based approach is losing ground. Mobile Network Operators (MNOs) are counting on Artificial Intelligence (AI) to transform wireless by increasing spectral efficiency, reducing signaling overhead, and enabling continuous network innovation through software upgrades. They may also be interested in new use cases like integrated sensing and communications (ISAC). All we need is an AI-native physical layer, so why not simply tailor the offline AI algorithms that have revolutionized image and natural language processing to the wireless domain? We argue that these algorithms rely on off-line training that is precluded by the sub-millisecond speeds at which the wireless interference environment changes. We present an alternative architecture, a universal neural receiver based on convolution, which governs transmit and receive signal processing of any signal in any part of the wireless spectrum. Our neural receiver is designed to invert convolution, and we separate the question of which convolution to invert from the actual deconvolution. The neural network that performs deconvolution is very simple, and we configure this network by setting weights based on domain knowledge. By \emph{telling} our neural network what we know, we avoid extensive \emph{offline training}. By developing a universal receiver, we hope to simplify discussions about the proper choice of waveform for different use cases in the international standards. Since the receiver architecture is largely independent of technologies introduced at the base station, we hope to increase the rate of innovation in wireless.

\end{abstract}
\IEEEpeerreviewmaketitle
\section{Introduction}
``AI and communication'' is one of the six key usage scenarios of IMT-2030~\cite{IMT-2030}.
Besides the communication aspect of requiring high area traffic capacity and user-experienced data rates to support distributed computing and AI applications, this usage scenario is also expected to include a set of new capabilities related to the integration of AI and compute functionalities into IMT-2030 as illustrated in the concept of AI-enabled cellular networks~\cite{shafin2020artificial}.
As a critical step, the 3rd Generation Partnership Project (3GPP) has initiated the exploration of AI in the 6G air interface~\cite{AIML6GR}.
This trend of standardizing and deploying AI for the air interface is anticipated to continue and evolve through 6G/NextG networks. 

The growing interests in this domain mainly arise from the intrinsic issues of \emph{network complexity}, \emph{model deficit}, and \emph{algorithm deficit} as detailed in~\cite{shafin2020artificial}, but tailored towards the air interface of the NextG mobile broadband networks. 
Specifically, the air interface of the NextG (e.g., 6G and beyond) is expected to be increasingly sophisticated with complex network topologies/numerologies, non-linear device components, and high-complexity processing algorithms.
Therefore, it becomes exceedingly challenging to utilize conventional model-based approaches in a scalable and efficient manner.   
Meanwhile, AI/ML-based data-driven approaches can effectively resolve these issues, providing an appealing alternative for the design of the NextG air interface.

Most of the introduced AI/ML-based strategies are focusing exclusively on tailoring the offline training that have revolutionized image and natural language processing to the NextG air interface~\cite{AIML6GR}.
However, very little success has been reported so far. 
Why?
One fundamental challenge is that 5G/NextG is a global technology, so models based on extensive offline training in New York City may disappoint when deployed in Delhi, or even in Dallas. A second fundamental challenge is machine learning at the Speed of Wireless. 
Is it even possible to learn in a sub-millisecond transmit time interval (TTI) when the propagation environment is changing very rapidly in the NextG air interface? 

In this paper, we develop machine learning methods that are inspired by and based on the traditional model-based approaches to learn at the Speed of Wireless in the NextG air interface.
We start from convolution, which governs the transmission and reception of any waveform in any part of the radio spectrum. 
By starting with a physical process that is common to all modes of wireless communication we are able to develop a universal receiver. 
Specifically, we present a neural receiver that is designed to invert convolution. 
We describe how it is able to implement online, real-time learning within each TTI without offline training for several physical layer (PHY) waveforms.
The key to achieving computational efficiency is to separate the question of which convolution to invert from the actual deconvolution. The neural network that performs deconvolution is very simple, and we configure this network by setting the weights of the underlying neural network based on domain knowledge. By \emph{telling} our neural network what we know, we avoid extensive offline \emph{training}. 


The organization of the paper is the following: Section~\ref{sec:SoW} will illustrate the dynamic nature of the radio environments in the air interface. 
The inherent challenges and potential solutions of applying AI/ML in the NextG air interface will be discussed in Section~\ref{sec:Learning}. 
The universal neural receiver will be introduced in Section~\ref{sec:UNR} with its basic principles, the geometric interpretation for its explainability as well as case studies for weight configuration for typical waveforms. Section~\ref{sec:ConOut} will contain the conclusion and the research outlook.
\section{The Speed of Wireless in the Air Interface}
\label{sec:SoW}
\begin{figure}[t] 
    \centering
    \includegraphics[width=1\linewidth]{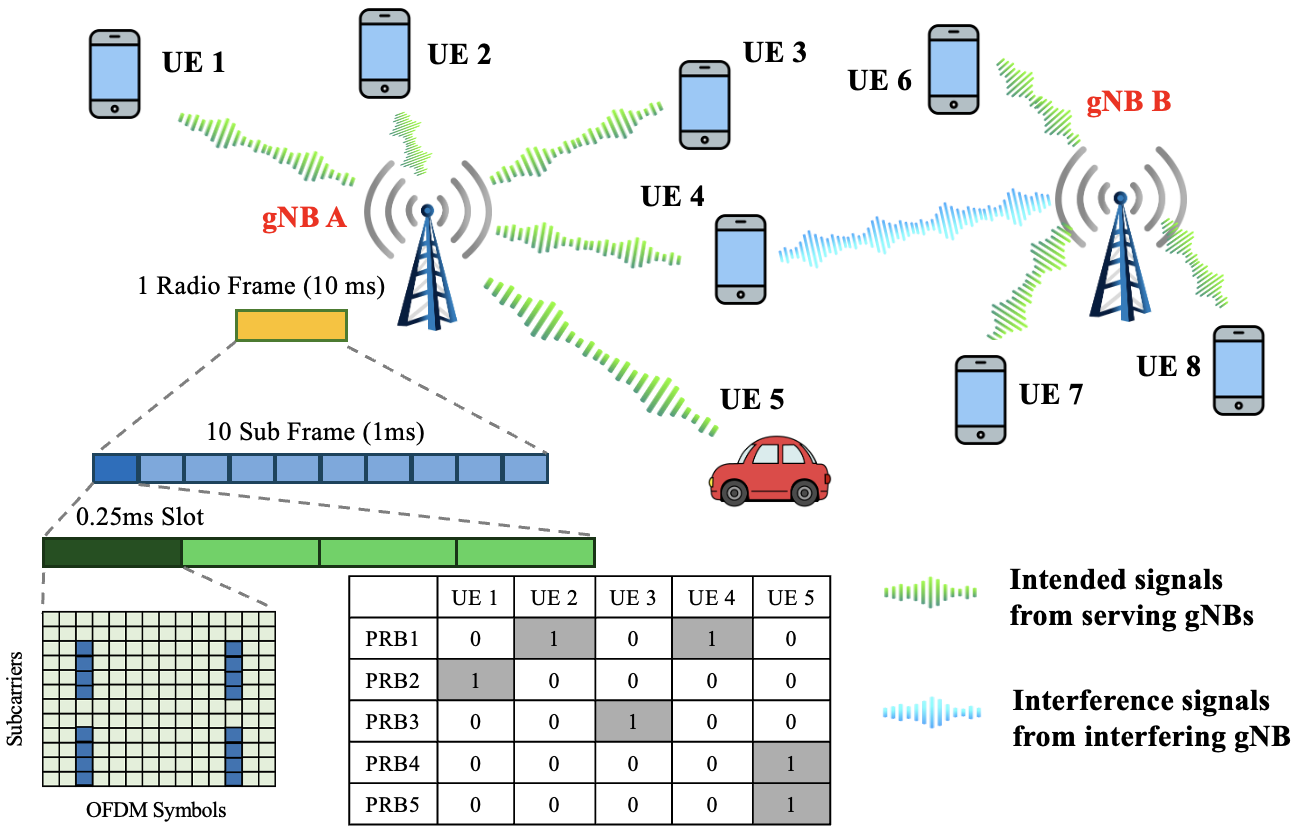}
    \caption{Speed of Wireless in the Air Interface.}
    \label{fig:SoW}
\end{figure}
The radio connections between mobile devices (termed UEs by 3GPP) and base stations (termed gNBs by 3GPP) define the air interface of a mobile broadband cellular network. 
If AI/ML is to transform NextG networks, then we need to meet the challenge of integrating AI/ML into the air interface, where interference changes on a sub-millisecond time scale.

Figure~\ref{fig:SoW} shows a current 5G/5G-Advanced air interface where $2$ gNBs serve $8$ UEs, with gNB~A serving $5$ UEs and gNB~B serving the remaining $3$ UEs. 
Data is partitioned into $10$ms radio frames for transmission over the radio link between gNB~A and UE $1$, and each radio frame is further divided into $10$ subframes each of duration $1$ms. Then, depending on the numerology, each subframe will be further partitioned into $1$, $2$, $4$, $8$, or $16$ slots with time duration ranging from $1$ms to $62.5\mu$s. 
The transmission time interval (TTI), comprising one or more slots, is the granularity at which the 5G/5G-Advanced air interface assigns resources -- subcarriers in the frequency domain and OFDM symbols in the time domain. 
The 6G/NextG air interface will be even more sophisticated.


\textbf{Scheduling Granularity and Complexity:} Resources are partitioned into physical resource blocks (PRBs) consisting of $12$ subcarriers (to provide frequency diversity) which the gNB assigns to a single user (single-user scheduling) or to multiple users (multi-user scheduling) during each TTI. There may be $100$s of active UEs within a cell, and the typical number of PRBs in a 5G/5G-Advanced air interface is between $50$ and $70$. 
A UE is assigned a block of PRBs, and the gNB schedules PRBs to provide smooth Quality of Service. The number of scheduling options is astronomical. 
For simplicity, consider the downlink in Figure~\ref{fig:SoW} where gNB~A has $5$ available PRBs to serve $5$ active UEs. 
For single-user scheduling, where each PRB is scheduled to at most one UE, there are $6^5$ possible scheduling options. 
For multiple-input multiple-output (MIMO) scheduling, gNB~A can schedule multiple users on a PRB. If we limit gNB~A to scheduling at most 2 UEs in each PRB, there are $(6 \times 5)^5$ possible options. 
Figure~\ref{fig:SoW} shows one possible scheduling option, but a great many more are possible, at each sub-millisecond TTI.

\textbf{Intrinsic Interference of Cellular Networks:} Cellular networks are interference-limited rather than noise-limited, and adaptation is designed mainly to counter the different sources of interference. 
5G/5G-Advanced networks typically reuse the same spectrum resources in every cell creating interference between adjacent cells. 
Multiple UEs that are scheduled on the same PRB will interfere with each other since it is not possible for the gNB to perfectly cancel multi-user interference~\cite{LiuMIMO}. 
For example, gNB A in Figure~\ref{fig:SoW} is serving both UE~$2$ and UE~$4$ in PRB $1$, while gNB~B is serving UE~$7$ in the same PRB. 
This results in intra-cell interference between UE~$2$ and UE~$4$, as well as inter-cell interference from UE~$7$. 
On top of that, doubly dispersive channels will introduce additional inter-subcarrier interference. 
In summary, the 5G/5G-Advanced air interface is already experiencing a myriad of complex interference scenarios, and this interference changes at a sub-millisecond granularity from TTI to TTI, as the gNB chooses from an astronomically large set of possible scheduling decisions. 
The NextG air interface is expected to be even more dynamic, as TTIs shrink and the number of active UEs grows to support new applications.

\textbf{Link and Rank Adaptation:} Link adaptation means that the modulation and coding scheme (MCS) will change to combat interference. 5G/5G-Advanced supports $32$ possible MCSs varying the constellation (from QPSK to 1024-QAM), code rate, and spectral efficiency. 
Rank adaptation means that the MIMO transmission rank will change from $1$ to $8$ to combat interference.
The challenge of adapting to interference at a TTI granularity in the 5G air interface is already intimidating. 
With NextG it is about to become more formidable. 

\section{Learning at the Speed of Wireless}
\label{sec:Learning}
\begin{figure}[t] 
    \centering
    \includegraphics[width=1\linewidth]{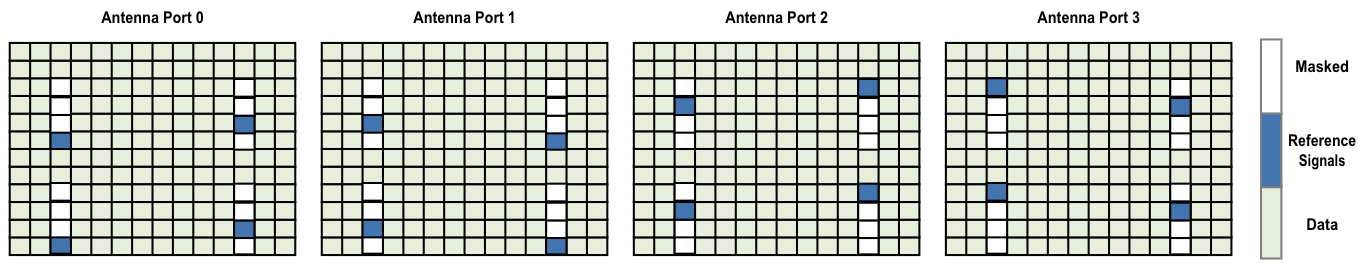}
    \caption{5G Reference Signal Pattern for $4 \times 4$ MIMO-OFDM.}
    \label{fig:RS}
\end{figure}
How do we migrate from \emph{model-based} signal processing to \emph{model-free} machine learning?
\begin{enumerate}
    \item \emph{Reference Signal-Based Channel Estimation in 5G/5G-Advanced}: Current systems use reference signals defined in the 5G/5G-Advanced standards~\cite{std3gpp38211} that vary with the communication scenario. Fig.~\ref{fig:RS} shows the reference signal pattern for a $4 \times 4$ MIMO-OFDM system in 5G/5G-Advanced. Intra-cell and inter-cell interference remain static over the TTI, since users are scheduled at the granularity of a TTI. The rank of MIMO-OFDM transmissions and the individual paths also remain constant. 
\end{enumerate}

5G receivers perform channel estimation based on knowledge of the reference signals. In the remainder of this paper, we will use reference signals and over-the-air (OTA) training samples interchangeably. After estimating the MIMO-OFDM channel for the reference signals, 2D interpolation methods are used to estimate the channel for an arbitrary resource element based on the time-averaged (empirical) channel statistics. 5G receivers then use the estimated MIMO-OFDM channel to perform joint detection of data symbols transmitted from multiple antennas. Given knowledge of modulation/coding and sufficiently many reference signals, the 5G receiver is able to adapt to interference that changes at TTI granularity. 

The current approach requires that we move back and forth between different mathematical models that govern signal propagation. 
As wireless networks become more and more complex, interest is growing in universal neural receivers that can operate model-free. 

\begin{enumerate}[resume]
    \item \emph{Uncertainty in Generalization:} The number of different interference scenarios within each TTI is very large, and we believe this makes it almost impossible for offline data to capture them all. Even if it were possible, the hybrid AI/ML models will find it difficult to use very limited over-the-air training data to adapt within a sub-millisecond TTI. 
\end{enumerate}

We briefly review potential sources of mismatch between the training dataset and subsequent signal detection:

\noindent\textbf{System Configuration Mismatch:} 5G/5G-Advanced features include the number of transmit/receive antennas, the number of slots within a subframe, and the subcarrier spacing. The number of possible combinations is very large – the number of antennas at a gNB can be $1$, $2$, $4$, $8$, $16$, the number of slots within a subframe can be $1$, $2$, $4$, $8$, $16$, and the subcarrier spacing can be $15$, $30$, $60$, $120$, or $240$ KHz depending on the carrier frequency.

\noindent\textbf{Channel Environment Mismatch:} The International Telecommunications Union~\cite{ITUEvaluation} distinguishes indoor from outdoor environments, urban from rural, as well as micro-cell from macro-cell.

\noindent\textbf{TTI-Based Transmission Adaptation Mismatch:} Transmission needs to adapt to a very large number of interference scenarios from TTI to TTI. For example, the rank of a $4 \times 4$ MIMO system can vary freely between $1$, $2$, $3$, and $4$, with the constellation varying from QPSK to $1024$-QAM.

It is possible to address uncertainty generalization by starting with an AI/ML model trained using a large corpus of offline data and to adapt in real time using the OTA training samples. 
A popular hybrid approach is model-agnostic meta-learning (MAML) where a small number of OTA training samples are used to adapt an initial model to new tasks in real-time. MAML is independent of the model architecture and has been shown to be effective at addressing channel environment mismatch. However, MAML is far from effective in addressing mismatch resulting from changes in interference scenarios from TTI to TTI. This is because the number of interference scenarios is very large, and the hybrid AI/ML model is biased towards the offline data, so that it is very difficult to use very limited OTA training samples to adapt within a sub-millisecond TTI.

On the one hand, by introducing additional OTA training samples we can address the issue of overfitting. 
On the other hand, reference signals/training samples constitute overhead since they do not carry data.

\begin{enumerate}[resume]
    \item \emph{Learning at the Speed of Wireless:} The number of reference signals is very limited, as illustrated for $4 \times 4$ MIMO-OFDM in Fig.~\ref{fig:RS}, where only $16$ out of 168 resource elements are available for over-the-air (OTA) training. There is a very large number of possible interference scenarios, but in every scenario the relationship between the transmitted and received signal is given by convolution. Signal detection inverts convolution, so we need to learn which convolution to invert, and the fundamental challenge is to do so at the Speed of Wireless. 
\end{enumerate}

We will show how to make effective use of the OTA training samples by incorporating domain knowledge into the design of the neural receiver. 
The neural network that performs deconvolution is very simple, and we configure this network by setting weights based on domain knowledge. 
By telling our neural network what we know, we avoid extensive offline training to achieve extreme efficient learning with sub-millisecond TTI.
In this way, we can achieve online real-time TTI-based training and testing to realize neural receiver that learns at the Speed of Wireless.

\section{Online Real-Time Universal Neural Receiver}
\label{sec:UNR}
There are a great many wireless channels while the number of OTA training samples is very limited. 
In this section, we design a neural receiver that is universal by taking advantage of the fact that all wireless channels are governed by convolution. 
Our receiver uses recurrent neural networks (RNNs) to implement deconvolution, where the network weights specify which deconvolution to implement. We describe how we combine domain knowledge and OTA training samples to set neural network weights. We conclude by comparing performance against conventional receivers, demonstrating significant gains for MIMO-OFDM and MIMO-OTFS at reduced complexity.

\subsection{Communication Over Inter-Symbol Interference Channels}
\begin{figure}
    \centering
    \includegraphics[width=0.8\linewidth]{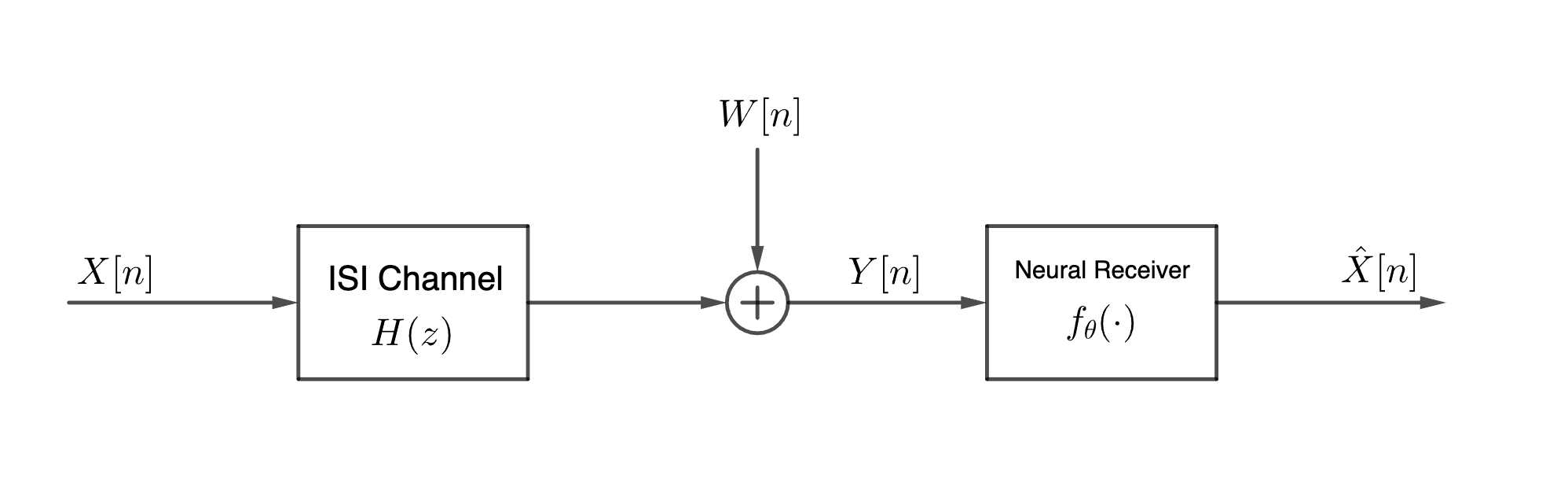}
    \caption{Neural Receiver for ISI Channel: neural network $f_\theta(\cdot)$ trained to recover $\hat{X}[n] = f_\theta(\{Y[n]\})$}
    \label{fig:neuralreceiver}
\end{figure}

Figure~\ref{fig:neuralreceiver} describes a wireless channel where the transmitted signal $X[n]$ is subject to inter-symbol interference (ISI) specified by a finite impulse response (FIR) filter $H(Z)$, then corrupted by additive noise $W[n]$ to yield a received signal $Y[n]$. We aim to design a parametrized family of neural networks $f_\theta(\cdot)$ that is able to approximate any filter $H(Z)$ and any noise distribution $W[n]$.   
We rely on OTA training to set the parameter $\theta$, so that after training $\hat{X}[n] = f_\theta(\{Y[n]\})$ provides an accurate estimate of $X[n]$. 

How then should we set the number of trainable parameters? The expressivity or \emph{model capacity}~\cite{Goodfellow-et-al-2016} of the neural network measures coverage of possible filters $H(z)$ and noise distributions $W[n]$. 
We can use a larger network with more trainable parameters, but this will require more OTA samples for training. 
Insufficient training samples will lead to larger generalization errors. 

As discussed in Section~\ref{sec:Learning}, the number of OTA training samples is tightly constrained in the NextG air interface. It is therefore very difficult to employ a large language model (LLM) as a neural receiver, because training and fine-tuning the very large number of weights would require orders of magnitude more OTA training samples. 
The number of OTA training samples limits what we can learn, and so the architecture of the neural network needs to capture what we know and do not need to learn. 

\subsection{Filtering and Deconvolution} 
We first develop intuition about receiver architecture by making the simplifying assumption that the noise $W[n]$ is additive white Gaussian.
As we relax this assumption and consider more realistic settings, we will discuss how domain knowledge informs the choices among possible architectures.

We write $H(Z) = h_0 + h_1 Z^{-1} + h_2 Z^{-2} + \ldots + h_k Z^{-k}$ and make the simplifying assumption that the inputs $\{X[n]\}$ are i.i.d. Gaussian.
We further assume that the noise power is small, so that the ideal neural receiver is the deconvolution filter $F^*(Z)$ that \emph{undoes} the channel, 
\begin{align}
F^*(Z) &= H^{-1}(Z) = \frac{1}{h_0 + h_1 Z^{-1} + h_2 Z^{-2} + \ldots + h_k Z^{-k}} \notag\\
&= \sum_{i=1}^k \;w_i \cdot \frac{1}{1- p_i Z^{-1}}.
\label{eqn:pfd}
\end{align}
The partial fraction decomposition given in \eqref{eqn:pfd} expresses $F^*(Z)$ as a weighted sum of first order infinite impulse response (IIR) filters. 
The \emph{weights} of $F^*(Z)$ are the parameters $w_i \in \mathbb C$, and the \emph{poles} of $F^*(Z)$ are the parameters $p_i \in \mathbb C$. 
We assume that all poles are within the unit circle, meaning that the ideal receiver is a causal stable filter. 

When we restrict to a scenario that is Gaussian and time-invariant, the optimal receiver $f_\theta(\cdot)$ is linear, parameterized either by $\theta = (h_0, \ldots, h_k)$ or by $\theta = \{(w_i, p_i), i=1, \ldots, k\}$. 
Either way, we know the structure of the optimal receiver, and we can discover the parameter $\theta$ through OTA training. 

\begin{figure}
     \centering
     \begin{subfigure}[b]{0.24\textwidth}
         \centering
         \includegraphics[width=\textwidth]{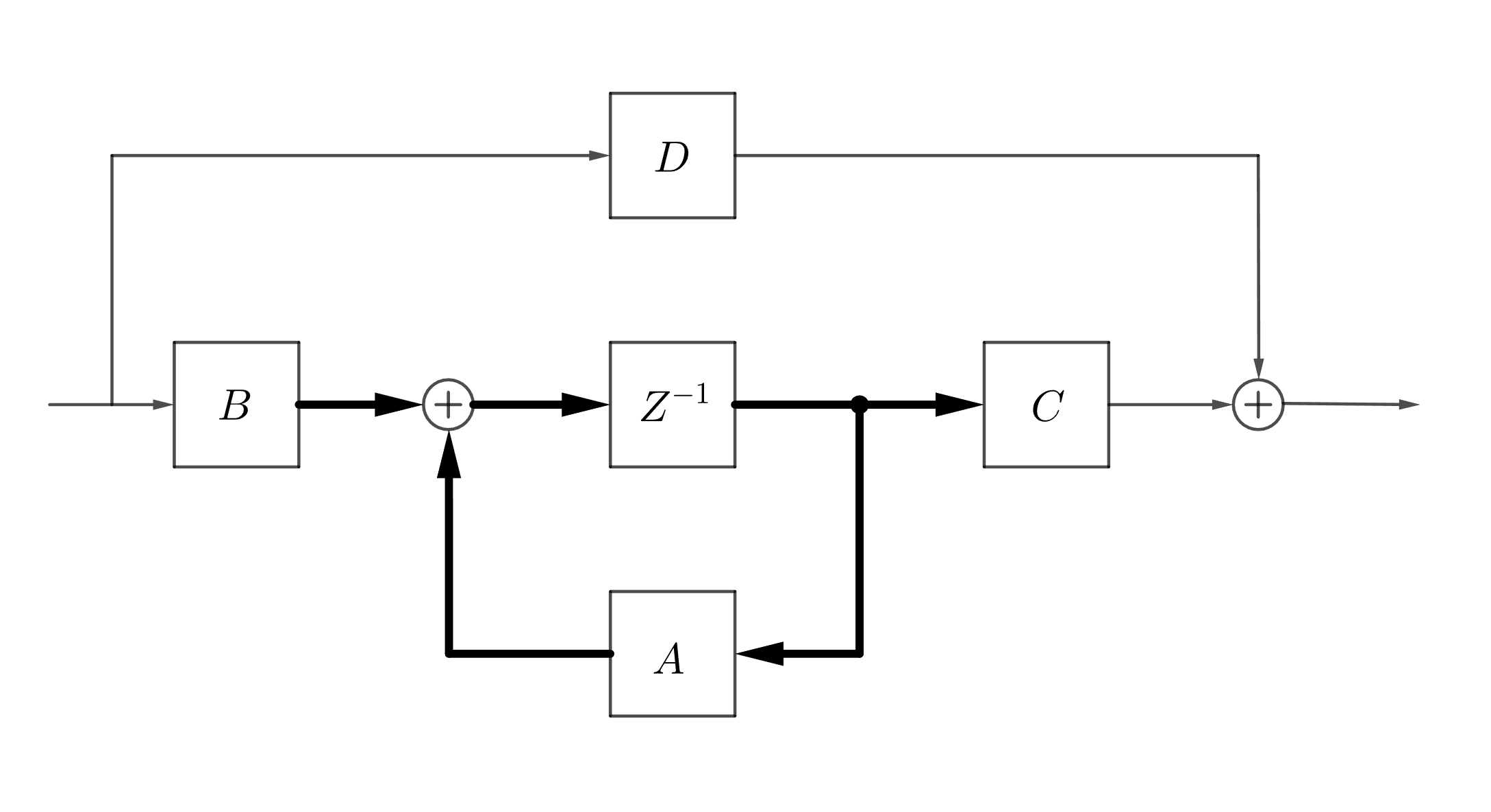}
         \caption{State Space Model}
         \label{fig:statespace}
     \end{subfigure}
     \hfill
     \begin{subfigure}[b]{0.24\textwidth}
         \centering
         \includegraphics[width=\textwidth]{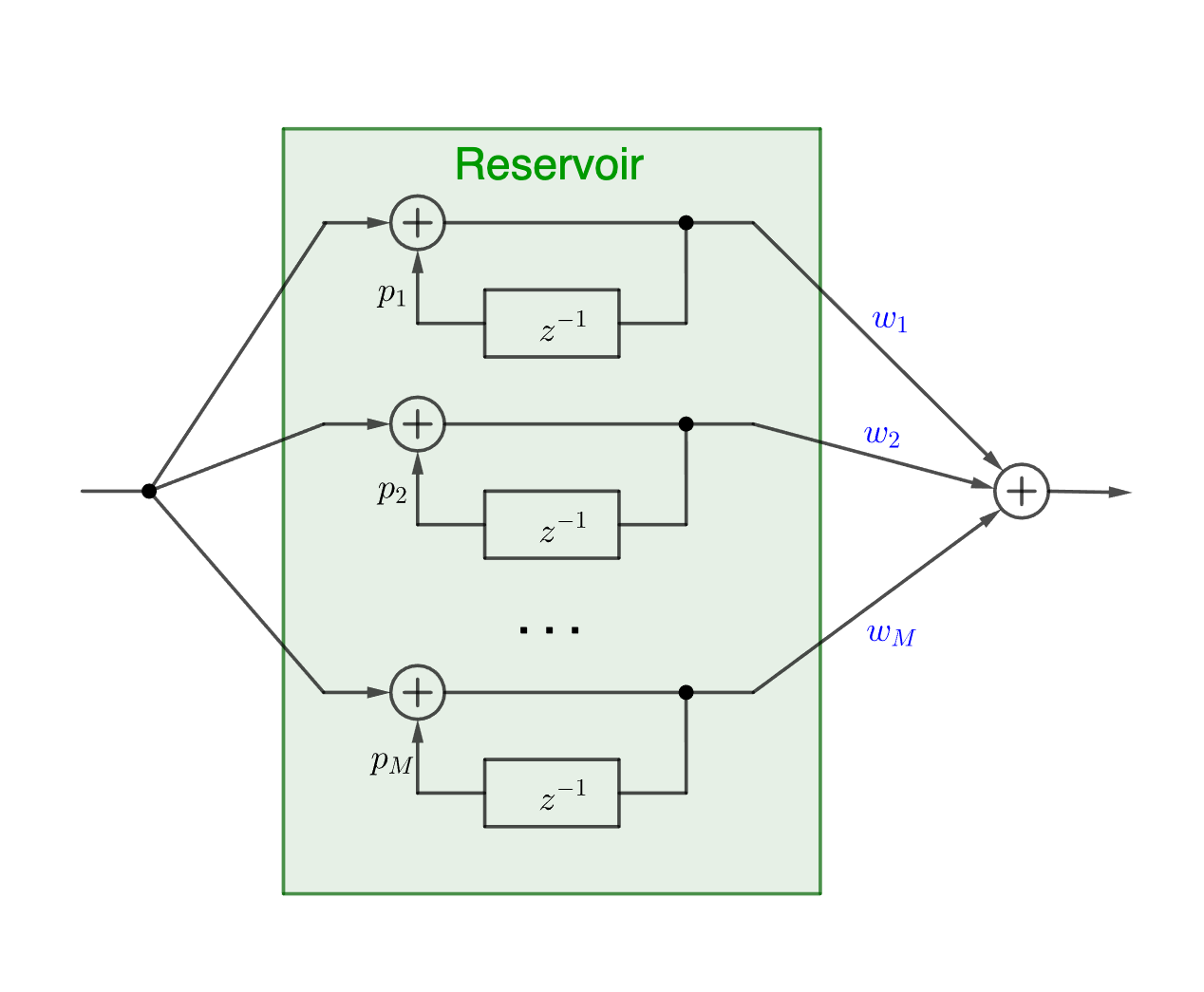}
         \caption{Parallel Reservoir}
         \label{fig:ESN}
     \end{subfigure}
        \caption{Two Ways to Implement the Deconvolution Filter}
        \label{fig:two_networks}
\end{figure}

\begin{figure}
    \centering
    \includegraphics[width=0.8\linewidth]{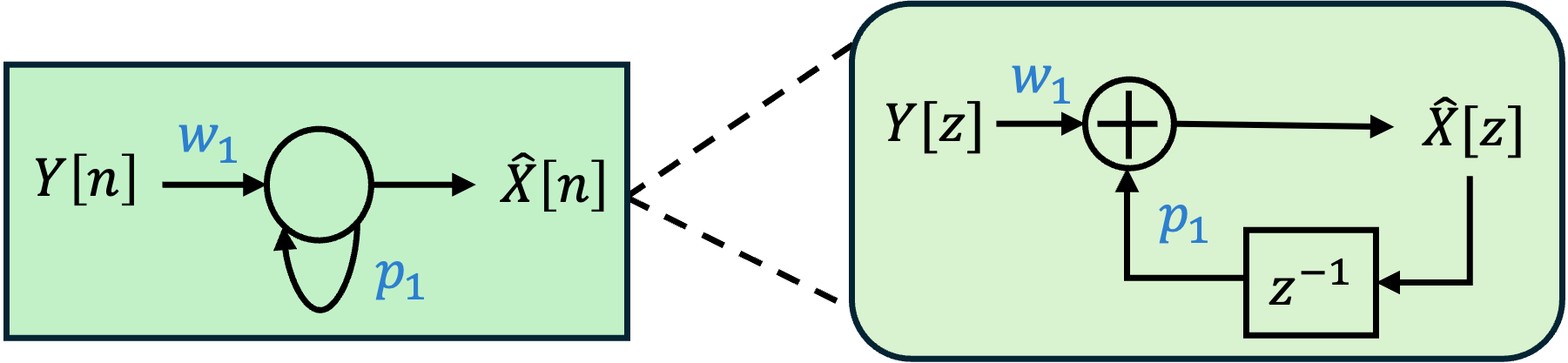}
    \caption{Single Linear Recurrent Neuron}
    \label{fig:1RNN}
\end{figure}

\subsection {Viewing Recurrent Neural Networks as Filters}
Among the different ways to represent the deconvolution filter, we focus on the state space model shown in Figure~\ref{fig:statespace} and the parallel reservoir shown in Figure~\ref{fig:ESN}. 
In Figure~\ref{fig:statespace}, we use thicker arrows to represent operations on $M$-dimensional state variables with feedback described by an $M \times M$ state transition matrix $A$.
We may view an eigenvector of $A$ as a realization of the single linear recurrent neuron shown in Figure~\ref{fig:1RNN}, with the corresponding eigenvalue specifying the pole of the IIR filter~\cite{ShashankxAI}.
This intuition connects the state sapce model with the parallel reservoir which we view as directly implementing~\eqref{eqn:pfd}, except that we use $M \ge k$ neurons rather than $k$ neurons.

We need to focus OTA training on those parameters that are easier to train. 
The impulse response $F^*(Z)$ is a linear combination of $k$ exponential sequences, each determined by a pole $p_i$.
If we were to know the poles $p_i$ it would be straightforward to determine the coefficients in the linear combination of exponential sequences, which are the weights $w_i$.
However, it may be much more difficult to learn the poles, since the input-output relation may not be equally sensitive to variations in different poles.
In fact, simple numerical experiments confirm that convergence can be very slow when gradient descent is used to learn poles from input-output sequences. This finding has a counterpart in the state space model, where it may be much more difficult to learn the state transition matrix $A$ than the parameters $B$, $C$, and $D$. 
Here numerical experiments show that convergence of (stochastic) gradient descent may be slow, and this can be confirmed by calculating the Hessian of commonly used loss functions with respect to $A$.

One solution is to avoid learning the poles. 
We provision $M > k$ neurons as shown in Figure~\ref{fig:ESN} with randomly chosen but fixed poles $p_i$, leaving only the weights $w_i$ as trainable parameters.
Intuitively, if there is a subset of neurons with poles that match or approximately match the $k$ desired poles, then it is possible to set the weights on these poles to the desired values in \eqref{eqn:pfd} and all other weights to be close to $0$, through the training process. 
The counterpart in the state space model is to add dimensions to the state variable and provision a randomly chosen state matrix $A$, leaving only $B$, $C$, and $D$ as trainable parameters. 
It is then possible to set some values of $B$, $C$, and $D$ to zero, and to select a subset of state variables with a dynamic that is close to that of the target filter $F^*(Z)$.

We rely on reservoir computing to provision a rich collection of dynamic subsystems, and train only the weights to select a linear combination that approximates the target filter. 
We have demonstrated in~\cite{ShashankxAI} that this is possible for the state space model and the parallel reservoir model.

\subsection{How to Configure RNN Weights}
Starting from the principle of reservoir computing with randomly generated poles, online real-time TTI-based neural receiver has been introduced in~\cite{RCMIMO1, RCMIMO2, RCMIMO3}. 
To further improve the learning efficiency, we tell our neural network what we know by configuring the weights of the underlying RNN so that we can focus OTA training on learning what we do not know. 
This organizing principle is the key to achieving online real-time TTI-based neural receiver. 

The first of four steps is to recognize that the learning objective is sensitive to the distribution of poles in the reservoir. 
When the target filter $F^*$ has a pole at $p$ and the reservoir has a few nearby poles, we can approximate the target pole with a linear combination of the nearby poles. 
If the target pole is close to the origin, this approximation will tend to cause a large $L_2$ error in the impulse response. 
We have explained in~\cite{SJere2024} why when designing a reservoir with $M$ poles, we seek a high density of poles close to the unit circle and a sparse density close to the origin. 
For a detailed quantitative analysis of how pole distribution/weight initialization affects convergence and performance, we refer the reader to~\cite{SJere2024}.
This first step is to recast distribution of interference environments in terms of distribution of poles in a reservoir.

The second step is to 
use the available domain knowledge to derive the statistical distribution of the channel parameters $h_i$. 
This domain knowledge can be either obtained from measurement-based channel models such as the 3GPP ones or from the time-averaged (empirical) channel measurements used in standard 5G receivers or from the combination of both.
Based on the available domain knowledge, we match the statistical distribution of poles in our reservoir to the statistical distribution of the channel parameters $h_i$. 
We have demonstrated very significant performance improvements in~\cite{ShashankxAI} from incorporating the time-averaged/empirical channel covariance in the configuration of the underlying RNN weights. 
It is also important to note that the input and recurrent weights of the underlying neural receiver can be entirely configured online based on time-averaged (empirical) channel statistics to enable output weight training/learning within each TTI. This mirrors the operation of conventional linear minimum mean square error (LMMSE)-based 5G receivers, wherein the filter weights are likewise derived from time-averaged channel statistics.

The third step is to modify our approach to accommodate channels $H(Z)$ that are described not by an FIR filter but by a more general rational function $H(Z) = a(Z) / b(Z)$. 
The deconvolution filter now takes the form
\begin{align}
    \label{eqn:skip}
    F^*(Z) = \sum_{i=1}^k \frac{w_i}{1-p Z^{-1}} + \sum_{j=0}^m \alpha_j Z^{-j}
\end{align}
and we need to account for the extra delay parameters $\alpha_j$.
We have explained in~\cite{Ummay} how to add skip connections to our RNN and include the $\alpha_j$ as trainable parameters. 
In this way, our receiver will be able to provide a comprehensive coverage of both IIR and FIR filers.

The fourth step is to modify our approach to accommodate non-linearities. 
Our discussion so far has focused on reconstructing a Gaussian input using a linear receiver subject to an $L_2$ loss function. 
In the NextG air interface, input symbols are drawn from a discrete constellation, and our aim is to minimize the probability of symbol detection error. 
One approach is to add a non-linear processing unit after the linear filter $f_\theta$ in Figure~\ref{fig:neuralreceiver}. 
We train $f_\theta$ to reverse linear mixing effects of the channel $H(Z)$, then rely on non-linear processing for symbol detection. 
This approach is realized by adding a non-linear unit at the output of Figure~\ref{fig:ESN} or by introducing recurrent neurons with non-linear activation functions.

\subsection{A Universal Neural Receiver} 
By configuring the weights that incorporate what we know, we are able to focus OTA training on learning what we do not know at the Speed of Wireless. 
Our neural receiver is able to adapt at TTI time scales to changes in MCS and MIMO rank.

\begin{figure}
     \centering
     \begin{subfigure}[b]{0.24\textwidth}
         \centering
         \includegraphics[width=\textwidth]{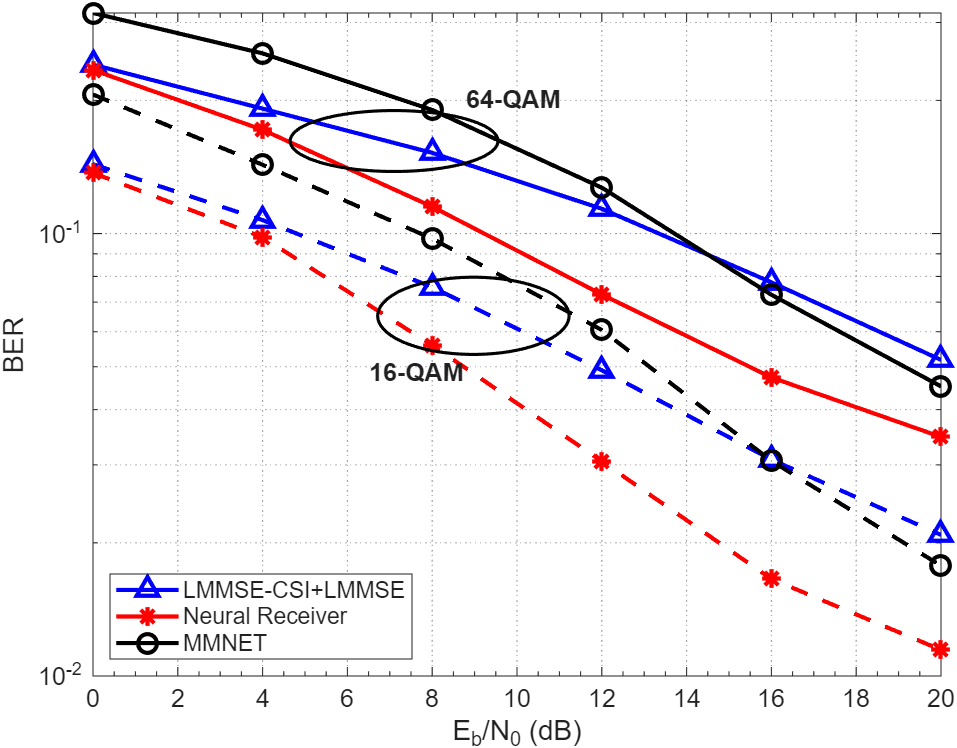}
         \caption{MIMO-OFDM}
         \label{fig:mimo_ofdm}
     \end{subfigure}
     \hfill
     \begin{subfigure}[b]{0.24\textwidth}
         \centering
         \includegraphics[width=\textwidth]{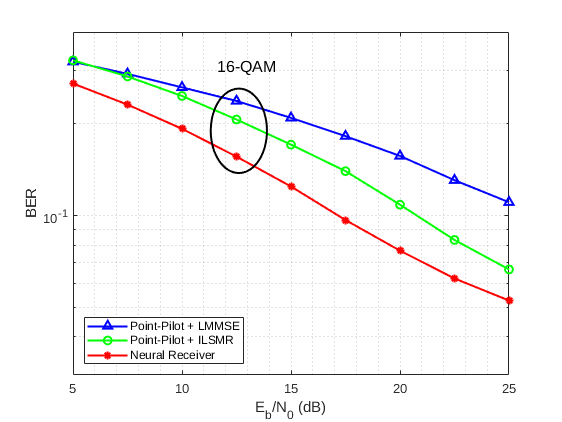}
         \caption{MIMO-MC-OTFS}
         \label{fig:mimo_otfs}
     \end{subfigure}
        \caption{BER comparisons between neural receiver and conventional methods for both MIMO-OFDM and MIMO-MC-OTFS.}
        \label{fig:compare}
\end{figure}
The design principle of first specifying a deconvolution problem, then performing the operation of deconvolution is universal. We see no fundamental obstacle to making it work for any frequency band and any waveform. We demonstrate our design principle is broadly applicable by applying it to MIMO-OFDM and to MIMO-OTFS (MC-OTFS) where the mathematics that governs convolution is different.
For MIMO-OFDM, we use the time-averaged (empirical) channel covariance matrix to configure RNN weights in the time-frequency (TF) domain, and for MIMO-OTFS we use the time-averaged (empirical) channel covariance to configure RNN weights in the delay-Doppler (DD) domain. 
Even though the neural receiver architecture is the same, the weight configuration are different for MIMO-OFDM and MIMO-OTFS respectively.

We conclude by evaluating the performance of our neural receiver for a $4 \times 4$ MIMO system using the 3GPP evaluation criteria. 
To be specific, we are using the 3GPP non-line-of-sight (NLoS) clustered delay line channel model-B (CDL-B) with $50$ns delay spread~\cite{3gpp_channel} transmitting $4$ data streams through the $4$ by $4$ MIMO system\footnote{Additional performance comparison of our neural receiver with conventional signal processing-based approaches in different channel environments and different system setups can be found in~\cite{ShashankxAI,RCMIMO2,RCMIMO3,Ummay}.}.
For both systems, we assume the same OTA reference signal overhead of $9.5\%$ (16 OFDM/OTFS symbols per TTI) as shown in the resource grid of Fig.~\ref{fig:SoW}.
Furthermore, we assume the carrier frequency is $3.8$ GHz, and we assume $256$ subcarriers with subcarrier spacing of $30$ kHz. 
The structure hyperparameters of our universal neural receiver are the same for both MIMO-OFDM and MIMO-MC-OTFS: the number of neurons in the input layer equals to the number of receive antenna ($4$), the number of neurons in the output layer equals to the number of data streams ($4$), and the number of neurons in the recurrent layer can be characterized in~\cite{ShashankxAI, Ummay} which is set to be $32$ in this case.
Both the input and recurrent weights of our neural receiver have been configured entirely online using time-averaged (empirical) channel statistics.
For MIMO-OFDM we set the user velocity to $30$ km/h, and for MIMO-OTFS we set it to be $450$ km/h.

Figure~\ref{fig:compare} compares performance of our neural receiver against conventional signal processing methods – the LMMSE receiver using LMMSE channel estimation based on 2D LMMSE interpolation for MIMO-OFDM as well as the LMMSE and iterative least square minimum residual (ILSMR)-based receivers for MIMO-MC-OTFS~\cite{ILSMR}. 
Figure~\ref{fig:mimo_ofdm} shows that for MIMO-OFDM with $16$-QAM and $64$-QAM, our neural receiver can achieve more than $2$dB gain over the LMMSE-based receiver while the other neural receiver based on the popular MMNet can not even catch up with the performance of the LMMSE-based receiver~\cite{RCMIMO3}. Note that the computational complexity of the LMMSE receiver for MIMO-OFDM is $\mathcal{O}(N_{sc}^3)$ while that of our neural receiver is $\mathcal{O}(N_{sc}+N_{rs}^2)$, where $N_{sc}$ is the number of subcarriers and $N_{rs}$ is the number of OTA reference signal resource elements per TTI. The computational complexity of the neural receiver based on MMNet is two orders of magnitude more than that of LMMSE~\cite{RCMIMO3}.
For a detailed quantitative analysis, we refer the reader to~\cite{RCMIMO3,ShashankxAI}.
From the performance evaluation shown in Figure~\ref{fig:mimo_ofdm} as well as the complexity analysis, it is evident that our neural receiver can achieve substantially improved detection performance even with lower complexity than the conventional LMMSE receiver.
Since the LMMSE receiver is conducting real-time TTI-based adaptive receive processing for 5G systems, this clearly demonstrates the online real-time nature of our neural receiver for the MIMO-OFDM waveform with extremely limited OTA reference signals.

Figure~\ref{fig:mimo_otfs} shows larger gains of neural receiver for MIMO-MC-OTFS with $16$-QAM over LMMSE, with gains of more than $2$dB over the much more complicated ILSMR receiver. 
Since there are no available neural receivers that are able to conduct online MIMO-OTFS detection, we can only compare our neural receiver with existing signal processing-based receivers. The computational complexity of the LMMSE receiver for MIMO-MC-OTFS can be expressed as $\mathcal{O}(N_{sc}^3N_{sym}^3N_{rx}^3)$, that of our neural receiver is $\mathcal{O}(N_{sc}+N_{rs}^2)$ while that of the ILSMR receiver is $\mathcal{O}(IKN_{sc}N_{rs}N_{rx}^2)$.
Here, $N_{sc}$ is the number of subcarriers (delay period), $N_{sym}$ is the number of symbols (Doppler period), $N_{rx}$ is the number of receiver antennas, $N_{rs}$ is the number of OTA reference signal resource elements including the guard band per TTI, $I$ and $K$ are the number of outer and inner iterations for ILSMR.
Typically, $IK > 100$ is chosen for improved performance. 
It can be seen from the complexity analysis that the complexity of LMMSE receiver for MIMO-MC-OTFS is much higher than its counterpart for MIMO-OFDM while the complexity of our neural receiver stays similar and is substantially smaller than that of the ILSMR receiver.
Similar to the case of MIMO-OFDM, it is evident that by telling the neural receiver what we know, we are able to learn at the Speed of Wireless for MIMO-MC-OTFS with a much lower computational complexity using the extremely limited OTA reference signals.

\section{Conclusion and Research Prospects}
\label{sec:ConOut}
AI/ML will need to learn at the Speed of Wireless in order to revolutionize the NextG air interface. We have presented a neural receiver architecture that is universal, since it is based on convolution which governs the relationship between transmit and received signals in any part of the wireless spectrum. Our architecture separates the question of which convolution to invert from the actual deconvolution; it expresses what we know in order to focus OTA training on what we still need to learn, and we have shown that this \emph{divide and conquer} approach enables online real-time TTI-based learning for receive processing. We have compared performance with conventional receivers, demonstrating significant gains for MIMO-OFDM and MIMO-MC-OFTS at reduced complexity. We suggest that the practicality of a low-complexity, real-time neural receiver that is agnostic to choice of waveform has the potential to greatly simplify standardization of wireless technology and increase the rate of innovation in wireless.

\bibliographystyle{IEEEtran}

\bibliography{IEEEabrv,reference.bib}
\end{document}